\documentclass[a4paper,11pt]{article}
\pdfoutput=1

\usepackage{amssymb,amsmath,bm}
\usepackage{a4wide}
\usepackage{color}
\usepackage{slashed}
\usepackage{graphicx}
\usepackage[latin1]{inputenc}
\usepackage{amsfonts}
\usepackage{lscape}
\usepackage{hyperref}
\usepackage{amsthm}
\usepackage{booktabs}
\usepackage{array}
\usepackage{rotating}
\usepackage[numbers, sort&compress]{natbib}
\usepackage{float}

\usepackage[small,bf]{caption}
\newcommand{\TeV}{{\, \rm TeV}}
\newcommand{\eps}{\epsilon}

\newcommand{\be}{\begin{equation}}
\newcommand{\ee}{\end{equation}}

\newcommand{\fref}[1]{Fig.~\ref{fig:#1}} 
\newcommand{\eref}[1]{Eq.~\eqref{eq:#1}}

\newcommand{\cref}[1]{Chapter~\ref{ch:.#1}}
\newcommand{\tref}[1]{Table~\ref{tab:#1}}

\newcommand{\nnl}{\nonumber \\}

\newcommand{\beq}{\begin{equation}} 
\newcommand{\eeq}{\end{equation}} 
\newcommand{\ba}{\begin{array}}  
\newcommand{\ea}{\end{array}} 
\newcommand{\bea}{\begin{eqnarray}}  
\newcommand{\eea}{\end{eqnarray} }  
\newcommand{\bal}{\begin{align}}
\newcommand{\eal}{\end{align}}   
\newcommand{\bi}{\begin{itemize}}  
\newcommand{\ei}{\end{itemize}}  
\newcommand{\ben}{\begin{enumerate}}  
\newcommand{\een}{\end{enumerate}}  
\newcommand{\bc}{\begin{center}}
\newcommand{\ec}{\end{center}} 
\newcommand{\bt}{\begin{table}}
\newcommand{\et}{\end{table}}  
\newcommand{\btb}{\begin{tabular}}
\newcommand{\etb}{\end{tabular}}

\newcommand{\cO}{{\mathcal O}}

\begin{document}

\vspace{1cm}
\begin{titlepage}
\vspace*{-1.0truecm}
\vspace{0.8truecm}

\begin{center}
\boldmath

{\Large\textbf{Lepton Flavor Non-Universality  in $B$-meson Decays   
   \\ from a $U(2)$ Flavor Model
}}
\unboldmath
\end{center}

\vspace{0.4truecm}

\begin{center}
{\bf Adam Falkowski$^a$, Marco Nardecchia$^b$, Robert Ziegler$^{c,d}$}
\vspace{0.4truecm}

{\footnotesize

$^a${\sl Laboratoire de Physique Th\'{e}orique, Bat.~210, Universit\'{e} Paris-Sud, 91405 Orsay, France \vspace{0.2truecm}}

$^b${\sl DAMTP, University of Cambridge, Wilberforce Road, Cambridge, CB3 0WA \vspace{0.2truecm}}

$^c${\sl Sorbonne Universit\'es, UPMC Univ Paris 06, UMR 7589, LPTHE, F-75005, Paris, France  \vspace{0.2truecm}}

$^d${\sl CNRS, UMR 7589, LPTHE, F-75005, Paris, France}

}
\end{center}

\begin{abstract}
\noindent 
We address the recent anomalies in semi-leptonic $B$-meson decays using a model of fermion masses based on the $U(2)$ flavor symmetry. 
The new contributions to  $b \to s \ell \ell$ transitions arise due to  a tree-level exchange of a $Z^\prime$ vector boson gauging a $U(1)$ subgroup of the flavor symmetry. 
They are controlled by a single parameter and are  approximately aligned to the Standard Model prediction, with constructive interference in the $e$-channel and destructive interference in the $\mu$-channel.   
The current experimental data on semi-leptonic $B$-meson decays can be very well reproduced without violating existing constraints from flavor violation in the quark and lepton sectors.
Our model will be tested by new measurements of $b \to s \ell \ell$ transitions and also by future electroweak precision tests, direct $Z^\prime$ searches, and $\mu$-$e$ conversion in nuclei. 
\end{abstract}

\end{titlepage}

\newpage

\renewcommand{\theequation}{\arabic{section}.\arabic{equation}}


\section{Introduction}
\setcounter{equation}{0}
While direct searches for new particles at the LHC have so far 
been inconclusive, recent results from the LHCb collaboration on semi-leptonic $B$-meson decays \cite{Aaij:2013qta,BKllLFULHCb,Aaij:2015esa} might provide the first indirect hint  of new physics beyond the Standard Model (SM). Starting with the $3 \sigma$ anomaly in $B \to K^* \mu^+ \mu^-$ angular observables \cite{Aaij:2013qta}, several other  observables involving $b \to s \ell \ell$ transitions have produced significant deviations from the SM predictions. The most  notable is the ratio of $B^{\pm} \to K^{\pm} \mu^+ \mu^-$ to $B^{\pm} \to K^{\pm} e^+ e^-$ branching ratios measured as $R_K = 0.745^{+0.097}_{-0.082}$  \cite{BKllLFULHCb} (deviating from the SM prediction by $2.6 \sigma$), as in this case the SM prediction $R_K \approx 1$ can be calculated with a very good  accuracy.  
The measured branching fraction of  $B_s \to \phi \mu^+ \mu^-$ \cite{Aaij:2015esa} is also low compared to the SM prediction. 

These anomalies could well be the result of statistical fluctuations, experimental problems, underestimated hadronic uncertainties, or  a combination of all three. 
Nevertheless, it is intriguing that many of the discrepancies can be simultaneously explained 
by assuming new physics contributions to the Wilson coefficients of 4-fermion operators with a b- and s-quark and 2 leptons~\cite{Descotes-Genon:2013wba,Altmannshofer:2013foa, Datta:2013kja, Beaujean:2013soa, Horgan:2013pva, Hurth:2013ssa,Alonso:2014csa,Hiller:2014yaa,Marco,HurthMahmoudiFit,AS2,Altmannshofer:2015sma}. 
Such contributions can be easily generated in explicit models through the exchange of a new $Z^\prime$ gauge boson~\cite{Altmannshofer:2014cfa,Crivellin:2015mga,Crivellin:2015lwa,Niehoff:2015bfa,Sierra:2015fma,CrivellinZ',Celis:2015ara,Greljo:2015mma,Belanger:2015nma,Niehoff:2015iaa,Altmannshofer:2015mqa} or leptoquarks~\cite{Hiller:2014yaa,Biswas:2014gga,GripaiosNardecchiaRenner,Sahoo:2015wya,Varzielas:2015iva,Becirevic:2015asa,Alonso:2015sja, Calibbi:2015kma, Sahoo:2015qha}. 
In order to address the $R_K$ anomaly,  lepton flavor universality has to be broken, which {\it generically} also implies lepton flavor violation (LFV)~\cite{Glashow:2014iga,Boucenna:2015raa,Varzielas:2015iva, Lee:2015qra, Guadagnoli:2015nra}. 
It is tempting to connect these patterns of flavor violation to the SM flavor puzzle, i.e. the experimentally  observed hierarchical structure of Yukawa couplings. 
Some papers in the literature have attempted to obtain this connection in the context of partial compositeness~\cite{GripaiosNardecchiaRenner,Niehoff:2015bfa,Niehoff:2015iaa} or leptoquark models~\cite{Varzielas:2015iva}. 
Other works~\cite{Crivellin:2015lwa,Celis:2015ara} considered gauged abelian symmetries that are  able to reproduce some features of the CKM matrix (but not the SM quark mass hierarchies). 
However, up to now, no model has been proposed that directly connects the anomalies to the generation of  fermion masses and mixings.

The purpose of this work is to provide a predictive model of this kind. We address the  anomalies in $b \to s \ell \ell$ transitions in the context of a light $Z'$ vector boson, whose couplings to fermions are governed by an underlying $U(2)_F$ symmetry that explains fermion masses and mixings. 
The original $U(2)$ models   proposed in the context of supersymmetry~\cite{BarbieriDvaliHall,BarbieriHallRomanino} have been disfavored by precision measurements in  the $B$-factories~\cite{RomaninoRoss}, as they predicted the relation $V_{ub}/V_{cb} = \sqrt{m_u/m_c}$ which was not borne out experimentally. 
However, it is not difficult to modify this prediction with a more general $U(1)_F$ charge assignment, as demonstrated in Ref.~\cite{DGPZ}. 

The model that we are presenting here is essentially a non-supersymmetric version of the one in Ref.~\cite{DGPZ}, in which the dominant source of deviations from the SM is due to the tree-level exchange of the $Z^\prime$ gauge boson associated to the $U(1)_F$  flavor group. Similarly to the supersymmetric model, the couplings of the $Z^\prime$ are approximately $U(2)_F$ symmetric, and flavor violating effects in the quark sector are suppressed by the small CKM mixing angles involving the 3rd generation.
 In contrast to Ref.~\cite{DGPZ}, we do not demand that the $U(1)_F$ charges are compatible with  $SU(5)$ grand unification. 
This generalization gives us  more freedom in the  charged lepton sector  to address the observed anomalies in $b \to s \ell \ell$ transitions. Once this is achieved, the parametric freedom in the model is to a large extent fixed by matching to the observed quark and charged lepton masses and quark mixing angles. 

In our scenario, the deviations in $b \to s \ell \ell$ observables arise  from a simple pattern of $Z^\prime$ contributions to the 4 relevant Wilson coefficients $C_{9,10}^{ee, \mu\mu}$. Namely, the new  contributions are aligned with the SM one (i.e. approximately left-handed) and controlled by a {\it single} parameter (the ratio of the $U(1)_F$ gauge coupling and the $Z^\prime$ mass multiplied by the  $b_L$-$d_L$ mixing angle) that sets its overall magnitude.
Moreover, they interfere constructively in the electron channel and destructively in the muon channel. As a consequence, we predict a simple pattern for the relevant $b \to s \ell \ell$ amplitudes: in the electron channels the SM predictions are rescaled by a factor $r_e >1$, whereas in the muon channels they are rescaled by  a correlated factor $r_\mu <1$. The current experimental data on $b \to s \ell \ell$ transitions determine the overall normalization of the $Z'$ contribution. This in turn fixes the predictions for other flavor-violating observables up to ${\cal O}(1)$ coefficients that span the parameter space of our model.  Comparing that with existing  constraints from $\Delta F =2$ and LFV observables,  we obtain bounds on these ${\cal O}(1)$ coefficients.  The strongest ones come from  $B_s$ and kaon mixing, electroweak precision measurement  in LEP-2, and, especially,  from $\mu$-$e$ conversion in nuclei.  These bounds disfavor large regions of the parameter space, but they nevertheless leave enough room to address the $B$-meson anomalies. The corollary is that our scenario will be decisively tested not only by upcoming new data from LHCb, but also from near future tests of LFV in $\mu \to 3 e$ decays and $\mu$-$e$ conversion in nuclei. Last but not least, if the $Z^\prime$ boson couples to fermions with electroweak strength,  it is within the kinematical reach of LHC.

This paper is organized as follows. In Section 2 we define the setup of the $U(2)_F$ flavor model and use its predictions for fermion masses and mixings to determine the couplings of the $Z^\prime$ gauge boson to fermions. In Section 3 we demonstrate that the resulting contributions from tree-level $Z^\prime$ exchange to Wilson coefficients controlling $b \to s \ell \ell$ transitions allow one  to address the $B$-meson anomalies. In Section 4 we study other constraints on the parameter space, and show that electroweak precision tests in LEP-2 and $\mu$-$e$ conversion in nuclei provide important constraints. We conclude in Section 5. In Appendix A we provide analytical  results for the eigenvalues and mixing angles of  the  quark Yukawa matrices.

\section{The Model}
\setcounter{equation}{0}

In this section we define our model with a $U(2)_F$ flavor symmetry. 
We first study its predictions concerning the fermion masses and mixings and demonstrate that the observed patterns in the quark and lepton sector can be reproduced. 
Then we discuss the physics of the $Z^\prime$ boson associated to the $U(1)_F$ factor of the flavor group. 
This  degree of freedom will be the origin of lepton flavor violation  in the $B$-meson sector that we discuss in the next section.

\subsection{Flavor Symmetries}

We first consider an extension of the SM with the global symmetry $U(2)_F \equiv SU(2)_F \times U(1)_F$ acting in the fermion's generation space. 
Here we restrict to the effective description   involving only SM fields and spurions parametrizing the breaking of $SU(2)_F \times U(1)_F$. 
We assume that the additional degrees of freedom needed to UV-complete this theory are heavy enough not to play role in the low-energy dynamics, i.e. the cutoff-scale $\Lambda$ of the effective theory is in the multi-TeV range. 
The first two  generations transform as a doublet under $SU(2)_F$, and the third generation is an  $SU(2)_F$ singlet. 
The $U(1)_F$ charges of all fermions are treated as free parameters for a while; they will be fixed later to reproduce the observed mass and mixing hierarchies.
The Higgs field is a total flavor singlet. 
The breaking of the flavor symmetry  is described by two scalar spurions:  $\phi$ transforming as ${\bf 2}_{X_{\phi}}$, and $\chi$ transforming as ${\bf 1}_{-1}$. 
These fields acquire the following vacuum expectation values (VEVs):
\begin{align}
\langle \phi \rangle & =  \begin{pmatrix} \eps_\phi \Lambda \\ 0 \end{pmatrix} \, , &
 \langle \chi \rangle & = \eps_\chi \Lambda \, ,
\end{align}
where we assume $\eps_{\phi, \chi} \ll 1$. 
We also define $\tilde{\phi} \equiv i \sigma^2 \phi^*$ which transforms as ${\bf 2}_{- X_{\phi}}$. 
In Table~\ref{tab:charges} we list the field content and their general transformation properties under the flavor group. In the next sections we will specify the $U(1)_F$ charges $X^F_i$ needed to reproduce fermion masses and mixings. 
\begin{table}[h]
\centering
\begin{tabular}{c|ccccc|ccccc|ccc}
& $Q_a$ & $U_a$ &$D_a$ &$L_a$ &$E_a$ &$Q_3$ & $U_3$ &$D_3$ & $L_3$ &$E_3$ &  $H$ & $\phi_a$ & $\chi$ \\
\hline
$SU(2)_F$ & ${\bf 2}$ & ${\bf 2}$ & ${\bf 2}$ & ${\bf 2}$ & ${\bf 2}$ &  ${\bf 1}$  & ${\bf 1}$ & ${\bf 1}$ & ${\bf 1}$ & ${\bf 1}$ & ${\bf 1}$ & ${\bf 2}$ & ${\bf 1}$ \\    
$U(1)_F$ & $X^Q_{1}$ & $X^U_{1}$ & $X^D_{1}$  & $X^L_{1}$  & $X^E_{1}$ & $X^Q_3 $ &$X^U_3 $ &$X^D_{3}$ &$X^L_{3}$ &$X^E_{3}$ &$0$ & $X_\phi$ & $-1$    
\end{tabular}
\caption{The field content and $U(2)_F$ quantum numbers. \label{tab:charges}}
\end{table}

As the fermions are in general charged under $U(2)_F$, Yukawa couplings require additional spurion insertions in order to be $U(2)_F$-invariant. 
This leads to non-renormalizable interaction  suppressed by appropriate powers of $\Lambda$. 
After inserting the spurion VEVs the cutoff dependence drops out, and Yukawa hierarchies arise from powers of the small parameters $\eps_{\phi, \chi}$. The resulting Yukawa matrices are of the form 
\begin{align}
\label{eq:yf}
y_f & \approx
\begin{pmatrix}
h_{11}^f \eps_\phi^2  \eps_\chi^{|X^{\cal D}_1 + X^{\cal S}_1 -2 X_\phi|} & h_{12}^f  \eps_\chi^{|X^{\cal D}_1 + X^{\cal S}_1 |} & h_{13}^f \eps_\phi  \eps_\chi^{|X^{\cal D}_1 + X^{\cal S}_3 - X_\phi|}  \\
- h_{12}^f  \eps_\chi^{|X^{\cal D}_1 + X^{\cal S}_1 |} & h_{22}^f \eps_\phi^2  \eps_\chi^{|X^{\cal D}_1 + X^{\cal S}_1 + 2 X_\phi|}& h_{23}^f \eps_\phi  \eps_\chi^{|X^{\cal D}_1 + X^{\cal S}_3 + X_\phi|}  \\ 
h_{31}^f \eps_\phi  \eps_\chi^{|X^{\cal D}_3 + X^{\cal S}_1 -  X_\phi|}  & h_{32}^f \eps_\phi  \eps_\chi^{|X^{\cal D}_3 + X^{\cal S}_1 + X_\phi|}  & h_{33}^f \eps_\chi^{|X^{\cal D}_3 + X^{\cal S}_3 |} 
\end{pmatrix} \, , 
\end{align}
where ${\cal D}=Q,L$, and ${\cal S}=U,D,E$. 
In each entry we omitted terms suppressed by more powers of $\epsilon_{\phi, \chi}$ coming from higher-dimensional terms in the effective theory.  
The absolute value appears because only positive powers of $\chi$ or $\chi^*$ are allowed in the effective Lagrangian.  
Note that, in contrast to the supersymmetric $U(2)$ model in Ref.~\cite{DGPZ}, there are no holomorphy constraints, which leads to a more general Yukawa pattern. 



We move to discussing the consequences of the Yukawa pattern in Eq.~(\ref{eq:yf}) for the fermion masses and mixing. 

\subsection{Quark Masses and Mixings}

In the quark sector, we fix $X^Q_3 = X^U_3 = 0$, so that the  top Yukawa coupling is not suppressed. 
Furthermore,  we impose the following constraints on the charges:  
\begin{align}
X_\phi &  < 0\, , &     X^Q_1 +  X_\phi &  \ge 0  \,  ,&   X^U_1 + X_\phi  & \ge 0 \, , &  X^D_3 &  \ge 0 \, ,  &    X^D_1 + X_\phi  & \ge 0\, . 
    \label{Qconditions}
\end{align}
With these constraints, we find the following up- and down-quark Yukawa matrices:
 \begin{align}
y_u & \approx 
\begin{pmatrix}
0 & h_{12}^u \eps^u_{12} &  0 \\
- h_{12}^u \eps^u_{12} & h_{22}^u \eps^u_{23} \eps^u_{32} & h_{23}^u \eps^u_{23} \\ 
 0 & h_{32}^u \eps^u_{32} & h_{33}^u  
\end{pmatrix} \, , 
&
y_d & \approx 
\begin{pmatrix}
0 & h_{12}^d \eps^u_{12} \frac{\eps^d_{32}}{\eps^u_{32}} &  0 \\
- h_{12}^d \eps^u_{12} \frac{\eps^d_{32}}{\eps^u_{32}} & h_{22}^d \eps^u_{23}   \eps^d_{32} & h_{23}^d \eps^u_{23} \eps^d_{33} \\ 
 0 & h_{32}^d \eps^d_{32} & h_{33}^d \eps^d_{33} 
\end{pmatrix} \, , 
\label{qyuks}
\end{align}
where we have defined
\begin{align}
\eps^u_{12} & \equiv \eps_\chi^{X^Q_1 + X^U_1} \, , &
\eps^u_{23} & \equiv \eps_\phi \eps_\chi^{X^Q_1 + X_\phi } \, , &
\eps^u_{32} & \equiv \eps_\phi \eps_\chi^{X^U_1 + X_\phi } \,  , & 
 \eps^d_{32} & \equiv \eps_\phi \eps_\chi^{ X^D_1 + X_\phi } \, , &
\eps^d_{33} & \equiv  \eps_\chi^{X^D_3  } \,  . 
\end{align}
The $y_{11}$, $y_{13}$, and $y_{31}$ entries are not exactly zero but one can show they yield subleading corrections to quark masses and mixings relatively suppressed at least by $\eps_\phi^2$. Thus, effectively, three texture zeros appear in the Yukawa matrix, much as in the supersymmetric models~\cite{DGPZ}.  
It was pointed out long ago \cite{BarbieriDvaliHall} that  the presence of these three texture zeros leads to relations among quark masses and mixings that work remarkably well from the phenomenological point of view. 

Yukawa matrices of the form in Eq.~(\ref{qyuks}) can be diagonalized in a fully analytic way. 
However,  it is more convenient  to first illustrate the most  important points in perturbative analysis. 
Indeed,  since mixing angles in the left-handed (LH) quark sector are known to be small, one can use them as the small parameter in which the eigenvalues and the remaining mixing angles are expanded. 
Ignoring ${\cal O}(1)$ coefficients, this gives the following rough estimates for the eigenvalues and the CKM matrix:
\begin{align}
y_t & \sim 1\, , &  y_c & \sim \eps^u_{23} \eps^u_{32} \, , &   y_u & \sim \frac{\eps^u_{12} \eps^u_{12} }{\eps^u_{23} \eps^u_{32}}  \, ,  \nonumber \\
 y_b & \sim \eps^d_{33} \, ,  &  y_s & \sim \eps^u_{23} \eps^d_{32}  \, , &  y_d & \sim  y_u \frac{\eps^d_{32}}{\eps^u_{32} }  ,  \nonumber \\
 V_{cb} & \sim  \eps^u_{23} \, , &  V_{us} & \sim  \frac{\eps^u_{12}}   {\eps^u_{23}  \eps^u_{32}} \, , &     V_{ub} & \sim  {\eps^u_{12} \over  \eps^u_{32}}    \, , &  \, .
\end{align}
We thus have 5 small parameters that set the order of magnitude of 8 observable quantities.  
Expressed in powers of the  Cabibbo angle $\lambda \approx 0.2$,  the magnitudes of the parameters consistent with experiment is 
\begin{equation}
\label{eq:epslambda}
\eps^u_{23} \sim \eps^u_{32} \sim \eps^d_{32}  \sim \eps^d_{33} \sim \lambda^2, \quad \eps^u_{12} \sim \lambda^5, 
\end{equation}
where the $y_s/y_c$ hierarchy must be explained by order 1 factors. 
The parametric size of rotation angles and matrices (see Appendix for our conventions) is then given by 
\begin{align}
\label{rotangleslambda}
s^{Lu}_{12} & \approx s^{Ru}_{12} \sim \lambda^2 \, ,  &
s^{Lu}_{13} &  \sim \lambda^4 \, , & s^{Ru}_{13} &  \sim \lambda^4 \, , & s_{23}^{Lu}  & \sim \lambda^2  \, , & s_{23}^{Ru}  & \sim \lambda^2 \, ,  \nonumber \\
s^{Ld}_{12} & \approx s^{Rd}_{12}  \sim \lambda \, , &
s^{Ld}_{13} & \sim \lambda^3 \, , & s^{Rd}_{13} & \sim    \lambda   \, , &
s_{23}^{Ld} & \sim \lambda^2 \, , & s_{23}^{Rd} & \sim 1\, 
\end{align}
and
\begin{align}
V^u_{L} \sim  V^u_{R} & \sim 
\begin{pmatrix}
1& \lambda^2 &  \lambda^4 \\
\lambda^2 & 1 & \lambda^2 \\
 \lambda^4  & \lambda^2   & 1
\end{pmatrix} \, , & 
V^d_{L} & \sim 
\begin{pmatrix}
1& \lambda &  \lambda^3 \\
\lambda & 1 & \lambda^2 \\
 \lambda^3  & \lambda^2   & 1
\end{pmatrix} \, , & 
V^d_{R} & \sim 
\begin{pmatrix}
1& \lambda &  \lambda \\
\lambda & 1 & 1 \\
 \lambda  & 1   & 1
\end{pmatrix} \, . 
\end{align}
The constraints in Eq.~(\ref{eq:epslambda}) are recovered (up to a mismatch in $\eps^u_{12}$ that again is ascribed to order one factors) 
when the $U(1)_F$ charges are fixed as 
\begin{align}
X^Q_3 = X^U_3 = 0, &   & X^Q_1 = X^U_1 = X^D_1 = X^D_3  = - X_\phi = 1\, , 
\label{eq:charges}
\end{align}
and the spurion VEVs are of the order $ \eps_\chi \lesssim  \eps_\phi \sim \lambda^2$. 
One robust conclusion is that, given the observed masses and mixings,  the $U(1)_F$ charges in the RH down sector should be universal. 
This has important  consequences for phenomenology, as we will discuss later on. 

We now improve on the above rough estimates, taking into account the ${\cal O}(1)$ coefficients.
For our purpose, it is convenient to express the observables in terms of physical quark masses and the unitary rotations that connect the flavor and mass basis.  
To this end we pick 4 rotation angles: $s_{23}^{Lu}$,  $s_{23}^{Ru}$, $s_{23}^{Ld}$, $s_{23}^{Rd}$.  
As  we show in Appendix, the remaining rotation angles, up to percent corrections, can be expressed in terms of these 4 angles and the quark mass ratios: 
\begin{align}
\label{eq:approxrotangles}
s^{Lu}_{12}  & \approx -s^{Ru}_{12} \approx  \sqrt{\frac{m_u}{m_c}}  \, , &
s^{Lu}_{13} & \approx - s^{Lu}_{23} s_{12}^{Lu} \, , & s^{Ru}_{13} & \approx    s^{Ru}_{23} s^{Lu}_{12} 
\, , \nonumber \\
s^{Ld}_{12}  & \approx -s^{Rd}_{12}  \approx  \sqrt{\frac{m_d}{m_s}}   \sqrt{c_{23}^{Rd}}  \, , &
s^{Ld}_{13} & \approx  - s^{Ld}_{23} s^{Ld}_{12}  \left( 1 - \frac{s^{Rd}_{23}}{c^{Rd}_{23} s^{Ld}_{23}} \frac{m_s}{m_b}\right) \, , & s^{Rd}_{13} & \approx     \frac{s^{Rd}_{23}}{c_{23}^{Rd}} s^{Ld}_{12} \, .
\end{align}
Using  Eq.~(\ref{eq:approxrotangles}),  the CKM elements up to phase factors can be expressed as:  
\beq
\label{eq:ckm}
V_{us} \approx   \sqrt{\frac{m_d}{m_s}}   \sqrt{c_{23}^{Rd}}, 
\quad 
V_{cb} \approx   s_{23}^{Ld} - s_{23}^{Lu}. 
\quad 
V_{ub} \approx    
\sqrt{\frac{m_u}{m_c}}   \left ( s_{23}^{Ld} - s_{23}^{Lu} \right) - 
\sqrt{\frac{m_d m_s}{m_b^2}}  \frac{s^{Rd}_{23}}{c_{23}^{Rd}}  .
\eeq 
In the original $U(2)$ models~\cite{BarbieriDvaliHall, BarbieriHallRomanino}   $s_{23}^{Rd}$  was taken to be small,  $s_{23}^{Rd}  \sim V_{cb}$. 
From \eref{ckm},  this  leads to the prediction $ |V_{ub}/V_{cb}|   \approx  \sqrt{m_u/m_c}$ which deviates from experimental data by more than $3 \sigma$. 
However, with a large RH $2$-$3$ rotation angle,  $s_{23}^{Rd} \sim c_{23}^{Rd}  \sim 1/\sqrt{2} $,  the  CKM angles can be well fit  (see also Refs.~\cite{RomaninoRoss, DGPZ}). 
The  other 3 rotation angles parametrizing the model can be  small, 
$s_{23}^{Lu}  \sim   s_{23}^{Ru}  \sim s_{23}^{Ld} \sim |V_{cb}|$.

One can explicitly verify that the charge assignment in \eref{charges} allows one to fit the masses and mixings in quark sector with coefficients $h^q_{ij}$ that are indeed ${\cal O}(1)$, and in turn check the validity of the above parametrization. 
For our fit, we take the masses and mixings calculated in the SM at  the scale 10~TeV \cite{AntuschMaurer}. 
With the Yukawa matrices 
\begin{align}
y_u & \approx
\begin{pmatrix}
0 & 4.9 \cdot \eps_\chi^2  & 0 \\
- 4.9 \cdot \eps_\chi^2 & 3.7 \cdot \eps_\phi^2  & 0.89 \cdot \eps_\phi \\  0 &1.3 \cdot \eps_\phi  & 0.79
\end{pmatrix} \, , 
&
y_d & \approx
\begin{pmatrix}
0 & 3.6 \cdot  \eps_\chi^2 & 0 \\ 
- 3.6 \cdot \eps_\chi^2 & - 0.62 \cdot  \eps_\phi^2 & 4.9 \cdot \eps_\phi \eps_\chi  \\  0 & - 0.20 \cdot \eps_\phi & 2.5 \cdot \eps_\chi
\end{pmatrix} \, ,
\end{align}
and the spurions VEVs 
\begin{align}
\eps_\chi & \approx 0.0040\, , & \eps_\phi & \approx 0.035 \, . 
\label{fitvevs}
\end{align}
one can reproduce the observed masses and mixings within the experimental errors. One can also check that the above Yukawa matrices give values for the $2$-$3$ mixing angles
\begin{align}
|s_{23}^{Lu}| & \approx 1.0 \cdot |V_{cb}| \, , & |s_{23}^{Ru}| & \approx 1.5 \cdot |V_{cb}| \, , &
|s_{23}^{Ld}| & \approx 2.1 \cdot |V_{cb}|  \, , & |s_{23}^{Rd}| & \approx 0.60 \, ,
\label{eq:benchmark}
\end{align}
Moreover the other rotation angles are in very good agreement with the approximate expression in Eq.~(\ref{eq:approxrotangles}). 
We will use Eq.~(\ref{eq:benchmark}) and Eq.~(\ref{fitvevs}) as a reference point for the natural values of the mixing angles in phenomenological analyses below. 

In summary, the quark masses and mixing angles can be successfully fit in our $U(2)_F$ flavor model with $\cO(1)$ Yukawa coefficients in the Lagrangian.  
The remaining freedom can be parametrized by four rotation angles. 
Their precise values depend on the  $\cO(1)$  coefficients,
however, barring large cancellations,  their order of magnitude is  fixed: 
$s_{23}^{Lu}  \sim   s_{23}^{Ru}  \sim s_{23}^{Ld} \sim |V_{cb}|$, and  $s_{23}^{Rd} \sim 1$.   

\subsection{Charged Lepton Masses}

In the lepton sector we focus on the charged lepton masses, and we ignore here the neutrino masses\footnote{This is mainly due to simplicity; there are no obvious obstacles to reproduce neutrino masses and mixings with Dirac neutrinos upon adding RH neutrinos.}. 
Therefore the rotation angles in the charged lepton sector are not constrained by phenomenology, which leaves more freedom in the choice of the model parameters. 
The simplest possibility is to take the  $U(1)_F$ lepton charges to be compatible with SU(5) grand unification, that is to say, the same as for the down-type quarks \cite{DGPZ}. 
However, one can show that such a choice  does not allow to address the lepton non-universality in $b \to s \ell \ell$ transitions, which is the primary goal in this paper. 
Therefore we make a different choice of the $U(1)_F$ charges: 
\begin{align}
\label{eq:u1f_lepton}
X_1^L &  = -3 -  X_1^E \, , & 
X_3^L & = 2 -  X_1^E \, , &
 X_3^E &  = 4 + X_1^E  \, , 
\end{align}
where $X_1^E$ does not enter into the lepton mass matrix and is left unspecified for the moment. 
This choice leads to the following lepton Yukawa matrix: 
\begin{align}
\label{eq:eyukX}
y_e & \approx
\begin{pmatrix}
h_{11}^e \eps_\phi^2 \eps_\chi  & h_{12}^e \eps_\chi^3 & h_{13}^e \eps_\phi \eps_\chi^2 \\
- h_{12}^e \eps_\chi^3&  0 & h_{23}^e \eps_\phi  \\   h_{31}^e \eps_\phi \eps_\chi^3 & h_{32}^e \eps_\phi \eps_\chi &   0
\end{pmatrix},  
\end{align}
where the  $2$-$2$ and $3$-$3$ diagonal elements are suppressed by $\eps_\phi^2 \eps_\chi^5$ and $\eps_\chi^6$, and thus can be neglected. 
The consequence is that the muon and tau Yukawa couplings are set by the $23$ and $32$ off-diagonal elements.
Indeed, diagonalizing \eref{eyukX} yields the Yukawa couplings  
\beq
y_e \approx h^e_{11}  \eps_\phi^2 \eps_\chi, \quad  y_\mu \approx \eps_\phi \eps_\chi h^e_{32}, \quad  y_\tau \approx h^e_{23} \eps_\phi. 
\eeq 
Using $\eps_{\phi,\chi}$ in Eq.~(\ref{fitvevs}), the correct lepton masses are recovered by fixing three $\cO(1)$ coefficients as 
$h^e_{11}  \approx 0.57, h^e_{23} \approx  0.29$, $h^e_{32}   \approx 4.3$. 
The rotation angles are then determined by the remaining ${\cal O}(1)$ coefficients: 
\begin{align}
s_{12}^{Le} & \approx  \frac{h^e_{13}}{h^e_{23}}\eps_\chi^2  \approx 5.5 \times 10^{-5} h^e_{13} \, , 
& s_{12}^{Re} & \sim \frac{h^e_{31} }{h^e_{32}}  \eps_\chi^2  \approx 4.2 \times 10^{-6}h^e_{31} \, ,
\nonumber \\
s_{23}^{Le} & \approx 1  \, , 
& s_{23}^{Re} & \approx 0\, ,  
\nonumber \\
 s_{13}^{Le} & \approx  \frac{h^e_{12} }{h^e_{32}}  {\eps_\chi^2 \over \eps_\phi} \approx  1.1 \times 10^{-4} h^e_{12}  \, , 
& s_{13}^{Re} & \approx - \frac{h^e_{12}}{h^e_{23}}  {\eps_\chi^3 \over \eps_\phi} \approx - 6.2 \times 10^{-6} h^e_{12}  \, .
\label{eq:se}
\end{align} 
In summary,  the charged lepton masses can be well reproduced with the $U(1)_F$ charge assignment in \eref{u1f_lepton}. 
The resulting structure of the Yukawa matrix in  \eref{eyukX}  leads to a large LH mixing between the 2nd and 3rd generation,  $s_{23}^{Le} \approx 1$,  and a small RH $2$-$3$ rotation,  $s_{23}^{Re} \approx 0$. 
The remaining freedom is  the charge $X^E_1$ and the three  $\cO(1)$ coefficients 
$h_{12}^e$, $h_{13}^e$, $h_{31}^e$ that set the magnitude of the $1$-$2$ and $1$-$3$ mixing angles. 
We will use this freedom later when addressing the $b\to s$ anomalies in a way that avoids phenomenological constraints. 

\subsection{$Z'$-Boson}

We now extend the model by promoting  $U(1)_F$ to a local symmetry  (as in Ref.~\cite{DGPZ}). 
We assume that the associated gauge boson is relatively light, with a mass in the TeV range. 
Note that the $U(1)_F$ symmetry without additional fermions is necessarily anomalous if the model explains fermion mass hierarchies.  
This is due to the relation~\cite{BinetruyLavignacRamond,BinetruyRamond, IbanezRoss}, 
\begin{align}
 \det y_u y_d  \sim  \eps_\chi^{4 X^Q_1 + 2 X^U_1 + 2 X^D_1 + 2X^Q_3 + X^U_3+ X^D_3}  \equiv \eps_\chi^{C_3} \, , 
\end{align}
where $C_3$ is the anomaly coefficient of the mixed $SU(3)^2 U(1)_F$ anomaly. 
As $U(1)_F$ is spontaneously broken by the VEVs of $\phi$ and $\chi$  at a scale $v^\prime \sim \eps \Lambda$, we assume that the anomaly is cancelled by unspecified dynamics (involving new chiral fermions) at the scale $\Lambda_{UV} \lesssim 4 \pi v^\prime$. 
Since the new gauge boson has a mass given by $M_{Z^\prime} = g^\prime v^\prime$,  
it can easily be the lightest new degree of freedom when $g^\prime$ is sufficiently small. We therefore ignore the additional heavy dynamics and concentrate on the effects of the $Z'$ gauge boson.
 
In the flavor basis, $Z^\prime$ couples to each fermion proportionally  to its $U(1)_F$ charge $X_i^a$, 
\begin{align}
{\cal L} & \supset g^\prime Z_\mu^\prime  \left[ X_i^Q  Q^\dagger_i  \overline{\sigma}^\mu Q_i    +  X_i^U  U^\dagger_i  \overline{\sigma}^\mu U_i +  X_i^D  D^\dagger_i  \overline{\sigma}^\mu D_i +    X_i^L  L^\dagger_i  \overline{\sigma}^\mu L_i +    X_i^E  E^\dagger_i  \overline{\sigma}^\mu E_i \right] . 
\end{align}
We have fixed these charges (except for $X_1^E$) to fit the observed fermion mass hierarchies. 
That fit also determines the unitary rotations that connect the flavor and the mass basis. 
Therefore, flavor non-universal effects mediated by $Z'$  are predicted in our model, up to an overall normalization determined by the $Z'$ mass and gauge coupling, and up to the  freedom  of choosing $X_1^E$ and order one Yukawa factors.  
In particular, the $SU(2)_F$ structure for the first two generations implies that flavor changing effects are entirely determined by the 3rd row of the rotation matrices. 
In the mass basis, the $Z'$ couplings take the form 
\begin{align}
{\cal L} & \supset g' \Delta^{f_i f_j}_{L,R} \, f_i^\dagger \overline{\sigma}^\mu f_j \, Z_\mu^\prime  \, ,
\end{align}
\begin{align}
\Delta_L^{f_i f_j} & = X^{\cal D}_1 \left[ \delta_{ij} - \frac{X^{\cal D}_1 - X^{\cal D}_3}{X^{\cal D}_1}(V_L^f)_{3i}(V_L^f)_{3j}^* \right]\, ,  & {\cal D} & = Q,L \, , \\
 \Delta_R^{f_i f_j} & =  X^{\cal S}_1 \left[ \delta_{ij} - \frac{X^{\cal S}_1 - X^{\cal S}_3}{X^{\cal S}_1}(V_R^f)^*_{3i}(V_R^f)_{3j} \right] \, , & {\cal S} & = U,D,E \, .
\end{align}
Note that flavor-violating couplings are proportional to the charge difference $X_1 - X_3$. 
As a consequence, with the charge assignments  in \eref{charges} there is no flavor violation in the RH down sector: $\Delta_R^{d_i d_j}  = \delta_{ij}$ . 
For the LH down quarks we find 
\beq
\Delta_L^{d_i d_j} 
\approx  \left ( \ba{ccc}
1  &  -s_{13}^{Ld}  s_{23}^{Ld}  &   s_{13}^{Ld} 
\\
-s_{13}^{Ld}  s_{23}^{Ld} & (c_{23}^{Ld})^2   & s_{23}^{Ld}
\\
 s_{13}^{Ld} &  s_{23}^{Ld}  &  (s_{23}^{Ld})^2 
\ea \right ) 
\sim  \left ( \ba{ccc}
1  &  \lambda^5 &   \lambda^3
\\
 \lambda^5 & 1   & \lambda^2
\\
  \lambda^3 &  \lambda^2  &   \lambda^4
\ea \right ) \, , 
\eeq 
Since $ s_{13}^{Ld}  \ll s_{23}^{Ld}$ (see \eref{approxrotangles}), 
the largest flavor violating effect of $Z'$ is in $b\to s$ quark transitions. 
This will be handy for addressing the recent $B$-meson anomalies, as we will discuss in the next section.  
For LH and RH up quarks we find
\beq
\Delta_X^{u_i u_j} 
\approx  \left ( \ba{ccc}
1  &  -s_{13}^{Xu}  s_{23}^{Xu}  &   s_{13}^{Xu} 
\\
-s_{13}^{Xu}  s_{23}^{Xu} & (c_{23}^{Xu})^2   & s_{23}^{Xu}
\\
 s_{13}^{Xu} &  s_{23}^{Xu}  &  (s_{23}^{Xu})^2 
\ea \right ) 
\sim  \left ( \ba{ccc}
1  &  \lambda^6 &   \lambda^4
\\
 \lambda^6 & 1   & \lambda^2
\\
  \lambda^4 &  \lambda^2  &   \lambda^4
\ea \right ) \, ,
\eeq 
where $X= \{ L,R \}$.

In the charged lepton sector the $Z^\prime$ couplings depend on the charge $X^E_1$ and to good approximation only on the $1$-$3$ rotation angles. 
\begin{align}
\label{eq:zee}
\Delta_L^{e_i e_j} & 
\approx  \left ( \ba{ccc}
- 3 - X_1^E &  5 s_{13}^{Le} &  0 
\\
5 s_{13}^{Le} & 2 - X_1^E & 0 
\\
0 & 0 & - 3 - X_1^E  
\ea \right ),   \quad 
 \Delta_R^{e_i e_j} 
 \approx   \left ( \ba{ccc}
X_1^E & 0  &   4 s_{13}^{Re}
\\
0 & X_1^E  & 0 
\\
4 s_{13}^{Re} & 0 & 4 +  X_1^E  
\ea \right )    \, .
 \nonumber  \\ 
\end{align}
The diagonal muon $Z'$ coupling is different from the electron and tau one at leading order, which is due to  the $2$-$3$ inversion in the LH sector. 
This feature of our model will allow us later to address the anomalies in $B$-meson decays involving muons and electrons. 
In the RH sector electrons and muons have approximately the same coupling to $Z'$ and the dominant flavor non-universal effects must involve the tau lepton. 
The largest flavor violating effects occur in the LH $\mu$-$e$ and RH $\tau$-$e$ transitions.  
 Note that the rotation angles setting the  magnitude of these lepton-flavor-violating effects are  fixed up to  ${\cal O}(1)$ factors and expected to be tiny, see \eref{se}. 
As a result,  lepton flavor violation in our model is suppressed at least by a factor of order $\eps_\chi^2/\eps_\phi \approx 5 \times 10^{-4}$ as compared to violation of lepton flavor universality.


\section{Phenomenology of $b \to s \ell \ell$ Transitions}
\setcounter{equation}{0}

We now turn to the predictions for  $b \to s \ell \ell$ transitions.
Our main goal is to address the recently observed violation of lepton flavor universality in $B$-meson decays \cite{BKllLFULHCb}. 
In our model, this anomaly is due to the exchange of a $U(1)_F$ $Z^\prime$ boson with mass in the multi-TeV range. 

Low-energy observables are controlled by the 4-fermion effective operators that arise from integrating out the $Z^\prime$ at tree level,
\begin{align}
\label{eq:4f}
{\cal L}_{\rm eff} & \supset  - \frac{g'{}^2}{2 M_{Z^\prime}^2} 
\left[  \Delta_{L}^{f_i f_j} \bar f_i \bar \sigma_\mu f_j + \Delta_{R}^{f_i f_j} f^c_i  \sigma_\mu \bar f^c_j \right ] 
\left[  \Delta_{L}^{f_k f_l} \bar f_k \bar \sigma_\mu f_l + \Delta_{R}^{f_k f_l} f^c_k  \sigma_\mu \bar f^c_l \right ]   \, . 
\end{align}
The 4-fermion operators in \eref{4f} include the ones relevant for $B \to K \ell \ell$ decays  which are customarily parametrized  by the following effective Hamiltonian  (see e.g. Ref.~\cite{CrivellinZ'}): 
\begin{align}
\label{eq:4fbs}
{\cal H}_{\rm eff}  \supset - {\alpha \over \pi } \frac{G_F}{\sqrt 2 } V_{tb} V_{ts}^* 
 \left[ \overline{s} \gamma_{\mu} P_L b \right]  
\left[ C_9^{\ell \ell'} \overline{\ell} \gamma^\mu  \ell'   + C_{10}^{\ell \ell'} \overline{\ell} \gamma^\mu  \gamma^5 \ell' \right] +{\rm h.c.} 
\end{align}
Note that the analogous 4-fermion operators with RH quarks are not generated in our model ($\tilde{C}_9^{\ell \ell'} = \tilde{C}_{10}^{\ell \ell'} = 0$), 
as a consequence of  the universal $U(1)_F$ charge assignment in the RH down sector.  
Matching \eref{4f} and \eref{4fbs},  the Wilson coefficients  are given by
\begin{align}
C_9^{\ell \ell'} & =   - \frac{ \pi g^{\prime 2}}{\sqrt 2 M_{Z^\prime}^2 G_F V_{tb} V_{ts}^* \alpha} 
\Delta_L^{sb} \left( \Delta_R^{\ell \ell'}  + \Delta_L^{\ell \ell'}  \right) \, ,  & 
 C_{10}^{\ell \ell'}  & =  
-  \frac{ \pi g^{\prime 2}}{\sqrt 2 M_{Z^\prime}^2 G_F V_{tb} V_{ts}^* \alpha}
  \Delta_L^{sb} \left( \Delta_R^{\ell \ell'}  - \Delta_L^{\ell \ell'} \right) \, . 
\end{align}
Focusing for now on the lepton flavor conserving operators with  electrons or muons, we have 
\begin{align} 
\label{eq:Cemu}
C_{9}^{ee}      & \approx  0.19  k     \, ,  &
 C_{10}^{ee}    & \approx  - 0.19  \, \left(1 + 2/3 X^E_1  \right) k   \, , \nonumber \\ 
C_9^{\mu \mu}   & \approx  - 0.13 k   \, ,  &
 C_{10}^{\mu \mu}  & \approx 0.13 \left(1 -  X^E_1  \right) k   \, , 
\end{align}
where we defined the parameter $k$ as 
\begin{align}
k \equiv 
\left(  \frac{20 \TeV}{M_{Z^\prime}/ g^\prime} \right)^2    \left( \frac{s_{23}^{Ld}}{|V_{cb}|} \right) \, , 
\label{eq:kdef}
\end{align}
and we used the numerical values 
$|V_{ts} | \approx |V_{cb} | \approx  0.041$, $\alpha(m_b) \approx 1/133$.

The explicit expressions for  the Wilson coefficients in \eref{Cemu} imply that large corrections to $B$-meson decays  involving muons are correlated with comparable corrections to the analogous observables involving electrons. 
While new physics in the muonic sector alone gives the most economical explanation of the LHCb anomalies (including $R_K$ \cite{Marco, AS2}), it has been emphasized that large corrections in  electron channels are not only allowed but could also (slightly) improve the goodness of the global fit \cite{HurthMahmoudiFit}.

\begin{figure}
\begin{center}
\includegraphics[scale=0.7]{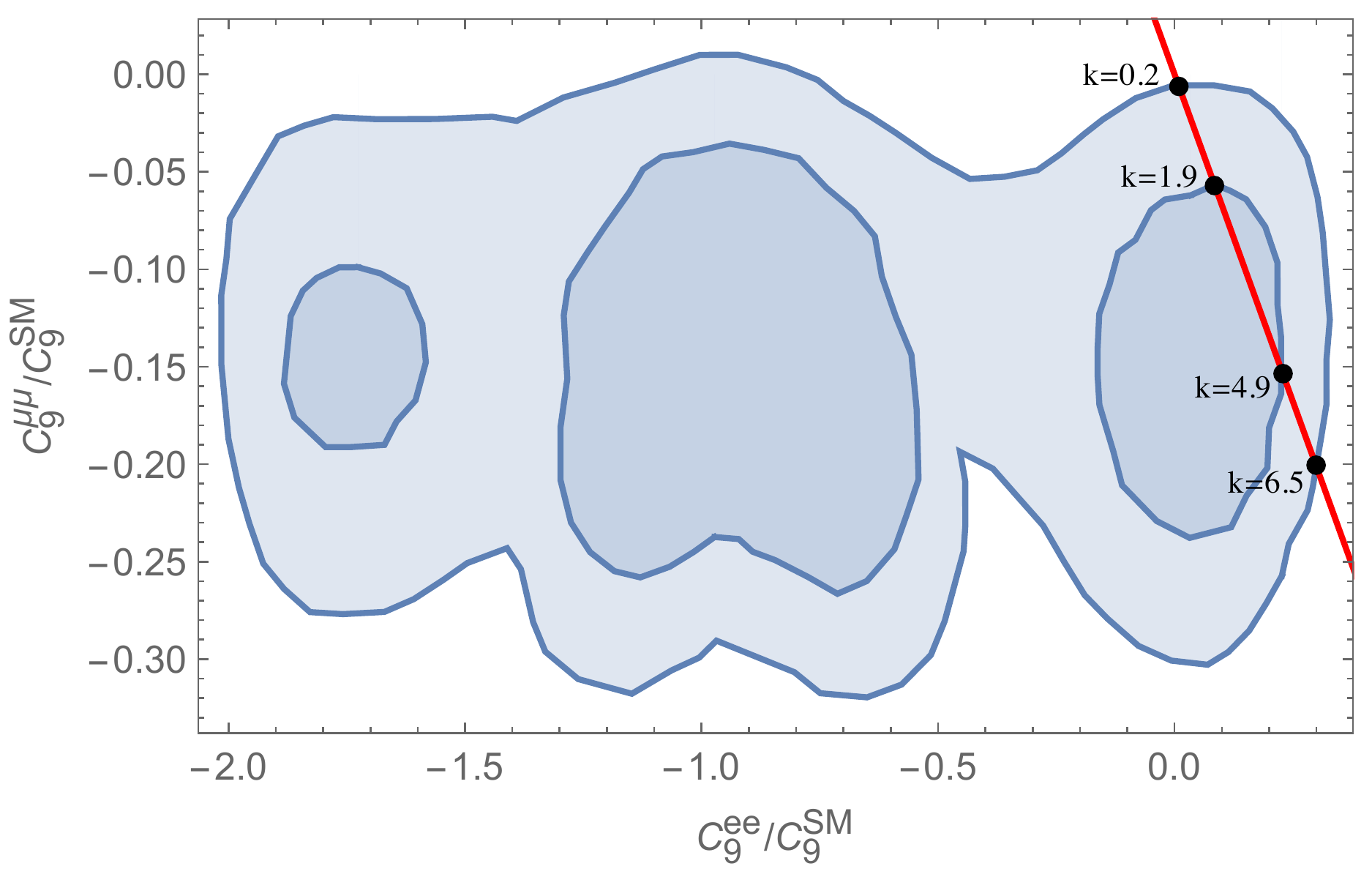}
\caption{Global fit results from \cite{HurthMahmoudiFit},  the blue (light blue) domain corresponds to the 1$\sigma$ (2$\sigma$) region. The red line is our model prediction for $X^E_1=0$ and varying $k$.}
\label{fig:LLfit}
\end{center}
\end{figure}

We can now identify the parameter space of our model where the measurements of semileptonic $b \to s \ell \ell$ transitions with $\ell=e,\mu$ are best reproduced. 
As can be seen from \eref{Cemu}, these observables depend on just 2 parameters:  
$k$ defined in \eref{kdef},  and the $U(1)_F $ charge $X^E_1$. 
For the moment,  we treat them as free parameters, various  precision constraints will be discussed in the next section.  
In order to find the best fit region for $k$ and $X^E_1$, we use the result of Ref.~\cite{HurthMahmoudiFit}.
The authors provide the results of the fit in the 2D planes  $(C_{9}^{e e},C_{9}^{\mu \mu})$, $(C_{10}^{e e},C_{10}^{\mu \mu})$, $(C_{9}^{\mu \mu},C_{10}^{\mu \mu})$ and $(C_{9}^{e e},C_{10}^{e e})$.
Ignoring possible correlations in the full 4D likelihood,  we identify the allowed range for $k$ for discrete  values of $X^E_1$ by requiring to simultaneously  remain inside the 68\% or 95\% confidence level regions in every 2D plane.
We obtain:
\begin{center}
\begin{tabular}{c | c | c}
\toprule
$X^E_1$ & $1\sigma$ Region & 2$\sigma$ Region \\
\midrule
\hline
-3 & - & $k \in [0.5,2.1]$ \\
-2 & - & $k \in [0.5,3.1]$ \\
-1 & - & $k \in [0.6,5.1]$ \\
0 & $k \in [2.7, 4.2]$ & $k \in [0.8,6.6]$ \\
1 & - & $k \in [1.2, 4.9]$ \\
2 & - & - \\
\hline
\bottomrule
\end{tabular}
\end{center}
This simplified analysis suggests that, while a reasonable fit to the $b \to s \ell \ell$  data is possible for a range of $X_1^E$,   the best case scenario is $X^E_1 = 0$. 
{\em In the rest of this paper we focus on that particular choice. } 
In this case, the new physics contributions mediated by $Z'$ are purely left-handed, 
 $C_{9}^{ee} = -C_{10}^{ee}$,  $C_{9}^{\mu \mu} = -C_{10}^{\mu \mu}$, 
 and we can derive constraints on $k$ using the 2-parameter fit of Ref.~\cite{HurthMahmoudiFit} for precisely this case. 
 This way, we find that the $1 \sigma$ confidence interval is $k \in [1.9,4.9]$ and the $2 \sigma$ one is  $k \in [0.2,6.5]$.
The allowed region of the parameters  $C_{9}^{ee}$-$C_{9}^{\mu \mu}$ parameter space overlaid with  the prediction of our model  is displayed in  Fig.~\ref{fig:LLfit}. 

We turn to discussing  predictions of our model. 
The case with  $X^E_1 =0$ is particularly simple because the SM contributions to the effective Hamiltonian  in \eref{4fbs} are also purely left-handed, 
 $C^{SM}_9 \approx -C^{SM}_{10}$, with $C^{SM}_9 (m_b) \approx 4.2$. 
Therefore the new physics contributions interfere constructively with the SM in the $e$-channel and destructively in the $\mu$-channel, resulting in a simple rescaling of $B$-meson decay rates by the factors $r_e$ and $r_\mu$  that are the same for all $b\to s ee$ and $b \to s \mu \mu$ processes\footnote{This a good approximation in the limit where contributions from fully-hadronic operators can be neglected.}, 
\beq
r_e \approx  (1+0.044 k)^2, \qquad r_\mu \approx  (1-0.029 k)^2  \, . 
\eeq 
For $k \in [1.9,4.9]$ we thus predict an enhancement in all the electron channels 
by $r_e  \in  [1.17,1.48]$, and a suppression in all the muon channels 
by $r_{\mu} \in  [0.73, 0.89]$. 
As a reference, in \tref{obs} we show the measured values and the SM predictions for various 
$b \to s \ell \ell$ observables. 
It is remarkable that most observables in the muon channel shows a deficit compared to the SM predictions, while the ones in the electron channels show some (albeit not statistically significant) enhancement. 

\begin{table}[tb]
\begin{center}
\begin{tabular}{c | c | c | c | c}
\toprule
Observable & $q^2$ $[\textrm{GeV}^2]$ & SM prediction & Measurement & Ratio \\
\midrule
\hline
$10^6 (\textrm{GeV})^2 \times \mathcal{B}(B^+ \to X_s e^+ e^-)$ & [1.0,6.0] &  $1.73 \pm 0.12$ & $1.93 \pm 0.55$ \cite{BXsllBaBar} & 1.12 $\pm 0.33$ \\
$10^6 (\textrm{GeV})^2 \times \mathcal{B}(B^+ \to X_s e^+ e^-)$ & [14.2,25.0] &  $0.20 \pm 0.06$ & $0.56 \pm 0.19$ \cite{BXsllBaBar} & 2.80 $\pm 1.27$ \\
\hline
$10^9 \times \mathcal{B} (B_s \to \mu^+ \mu^-)$ & - &  $3.54 \pm 0.27$ & $2.9 \pm 0.7$ \cite{BsmumuLHCb, BsmumuCMS} & 0.8 $\pm 0.2$ \\
$10^9 (\textrm{GeV})^2 \langle \frac{d \mathcal{B}}{d q^2} \rangle (B^0 \to K^0 \mu^+ \mu^-)$ & [1.1,6.0] &  $31.7 \pm 9.4$ & $18.7 \pm 3.6$ \cite{BKmumuLHCb} & 0.59 $\pm 0.21$ \\
$10^9 (\textrm{GeV})^2 \langle \frac{d \mathcal{B}}{d q^2} \rangle (B^0 \to K^0 \mu^+ \mu^-)$ & [15.0,22.0] &  $13.6 \pm 2.0$ & $9.5 \pm 1.7$ \cite{BKmumuLHCb} & 0.70 $\pm 0.16$ \\
$10^9 (\textrm{GeV})^2 \langle \frac{d \mathcal{B}}{d q^2} \rangle (B^+ \to K^+ \mu^+ \mu^-)$ & [1.1,6.0] &  $34.8 \pm 10.3$ & $24.2 \pm 1.4$ \cite{BKmumuLHCb} & 0.70 $\pm 0.21$ \\
$10^9 (\textrm{GeV})^2 \langle \frac{d \mathcal{B}}{d q^2} \rangle (B^+ \to K^+ \mu^+ \mu^-)$ & [15.0,22.0] &  $14.8 \pm 2.0$ & $12.1 \pm 0.7$ \cite{BKmumuLHCb} & 0.82 $\pm 0.12$ \\
$10^9 (\textrm{GeV})^2 \langle \frac{d \mathcal{B}}{d q^2} \rangle (B^+ \to K^* \mu^+ \mu^-)$ & [1.1,6.0] &  $50.5 \pm 28.6$ & $36.6 \pm 8.7$ \cite{BKmumuLHCb} & 0.72 $\pm 0.45$ \\
$10^9 (\textrm{GeV})^2 \langle \frac{d \mathcal{B}}{d q^2} \rangle (B^+ \to K^* \mu^+ \mu^-)$ & [15.0,19.0] &  $61.5 \pm 34.8$ & $39.5 \pm 8.5$ \cite{BKmumuLHCb} & 0.64 $\pm 0.39$ \\
$10^6 (\textrm{GeV})^2 \times \mathcal{B}(B^+ \to X_s \mu^+ \mu^-)$ & [1.0,6.0] &  $1.66 \pm 0.12$ & $0.66 \pm 0.88$ \cite{BXsllBaBar} & 0.40 $\pm 0.53$ \\
$10^6 (\textrm{GeV})^2 \times \mathcal{B}(B^+ \to X_s \mu^+ \mu^-)$ & [14.2,25.0] &  $0.24 \pm 0.07$ & $0.60 \pm 0.31$ \cite{BXsllBaBar} & 2.50 $\pm 1.48$ \\
\hline
\bottomrule
\end{tabular}
\caption{\label{tab:obs} 
The measured values and the SM predictions for various $b \to s \ell \ell$ observables used in the fit of Ref.~\cite{HurthMahmoudiFit}. 
}
\end{center}
\end{table}

Lepton flavor universality is often tested by measuring ratios of branching fractions of semileptonic $B$-meson decays.  In our model we have 
\begin{equation}
R_{X} = \frac{\mathcal{B} (B \to X \mu^+ \mu^-  )}{\mathcal{B}  (B \to X e^+ e^-)}\approx \frac{r_{\mu}}{r_{e}} = \left( \frac{1-0.029 k}{1+0.044 k} \right)^2 \, , 
\end{equation}
where $X=K,K^*,\phi,X_s$. 
The interval  $k \in [1.9,4.9]$  corresponds  $R_{X} \in [0.50,0.76]$, 
which should be compared to the LHCb measurement $R_K = 0.745^{+0.097}_{-0.082}$. 
Future improvements in the precision of $R_K$ and other measurements will be crucial for testing our model, since we predict  a rather low value for these observable. This is actually supported by measurements of inclusive $B \to X_s \ell \ell $ decay ratios from BaBar ($R_{X_s} = 0.58 \pm 0.19 $)~\cite{BXsllBaBar} and Belle ($R_{X_s} = 0.42 \pm 0.25 $)~\cite{BelleXs}, although with large errors. Another test of lepton-non-universality is provided by double ratios \cite{HillerSchmaltzDiagnosis} such as $R_{K^*}/R_{K}$.  
 In our model, as in any scenario with $\tilde{C}_{9,10} = 0$, all these double ratios are predicted to be equal to one.

\begin{table}[tb]
\begin{center}
\begingroup
\begin{center}
\begin{tabular}{c | c | c | c }
\toprule
Decay & Branching ratio & Ref. & Type \\
\midrule
\hline
$B^+ \to K^+ \mu^+ \mu^-$ & $\left( 4.29 \pm 0.22 \right) \times 10^{-9}$ & \cite{BKllLFULHCb} & LFC \\
$B^+ \to K^{+} \mu^{\pm} \tau^{\mp} $ & $ < 4.8  \times 10^{-5}$ & \cite{PDG2014}  & LFV \\
$B^+ \to K^{+} e^{\pm} \tau^{\mp} $ & $ < 3.0  \times 10^{-5}$ & \cite{PDG2014}  & LFV \\
$B^+ \to K^{+} e^{\pm} \mu^{\mp} $ & $ < 9.1 \times 10^{-8}$ & \cite{PDG2014}  & LFV \\
\hline
$B_s \to \mu^+ \mu^-$ & $\left( 2.9 \pm 0.7 \right) \times 10^{-9}$ & \cite{BsmumuLHCb, BsmumuCMS} & LFC \\
$B_s \to e^{\pm} \mu^{\mp}$ & $ < 1.1 \times 10^{-8}$ & \cite{BemuLHCb}  & LFV \\
\hline
\bottomrule
\end{tabular}
\end{center}
\endgroup
\end{center}
\caption{\label{TableBR} Experimental constraints on the branching fraction of several  lepton flavor conserving (LFC) and lepton flavor violating (LFV) $B$-meson decays. }
\end{table}
Finally, we  comment on the predictions concerning LFV $B$-meson decays. 
The Wilson coefficients of 4-fermion operators mediating these decays are suppressed by additional powers of small parameters, e.g. 
\beq
C_{9}^{\mu e} \approx - C_{10}^{\mu e} \approx 
-0.3  \, s^{Le}_{13} \, k \, , 
\eeq 
From \eref{se}, typical values of the lepton mixing angle are $s^{Le}_{13} \sim 10^{-3}$-$10^{-5}$. 
This implies the rate of lepton flavor violating decays is suppressed by at least 6 orders of magnitude compared to the lepton flavor conserving ones. Given the present sensitivity summarized in Table~\ref{TableBR}, this will not be observable in the near future. LFV decays involving tau leptons are even more suppressed.  

\section{Constraints from Flavor Violation and Direct Searches}
\setcounter{equation}{0}

In this section we discuss constraints on the parameters of our model from 
 $\Delta F = 2$ flavor transitions, LFV decays of leptons, LEP-1 and LEP-2 electroweak precision observables,  and direct $Z^\prime$ searches at the LHC. 
 We will show that it is possible to address the observed violation of  lepton-flavor universality in $B$-meson decays without violating these constraints.  

\subsection{Constraints from $\Delta F = 2$ Observables}

The $Z'$ boson exchange generates 4-quark operators mediating $\Delta F = 2$ transitions. 
In  the notation of e.g. Ref.~\cite{KboundsETM} these are denoted $O_1$ (with 4 LH quarks), $\tilde{O}_1$ (with 4 RH quarks), and $O_5$ (with 2 LH and 2 RH  quarks). 
Since $\Delta_R^{d_i d_j}  = 0$, for the down-type quark only $O_1$  is generated. 
Their (in general complex) Wilson coefficients are given by 
\begin{align}
C_1^K & = \frac{g'{}^2 \Delta_L^{sd} \Delta_L^{sd}}{2 M_{Z^\prime}^2} \, , &
C_1^{B_d} & = \frac{g'{}^2 \Delta_L^{bd} \Delta_L^{bd}}{2 M_{Z^\prime}^2} \, , &
C_1^{B_s} & = \frac{g'{}^2 \Delta_L^{bs} \Delta_L^{bs}}{2 M_{Z^\prime}^2} \, , 
\end{align}
and numerically one has (using $m_d/m_s \approx 0.05$)
\begin{gather}
C_1^K  =  \frac{1.8 \times 10^{-10}}{\TeV^2}
\left( \frac{s^{Ld}_{23}}{|V_{cb}|} \right)^3 c_{23}^{Rd} \, k   \, , \\
C_1^{B_d}  = \frac{1.1 \times 10^{-7}}{\TeV^2}  
\left( \frac{s^{Ld}_{23}}{|V_{cb}|} \right) c_{23}^{Rd} \, k   \, , \qquad
C_1^{B_s}  = \frac{2.1 \times 10^{-6}}{\TeV^2}   \left( \frac{s^{Ld}_{23}}{|V_{cb}|} \right) \, k    \, ,
\end{gather}
where $k$ is defined in \eref{kdef} and it needs to be $O(1)$ for the model to address the  $B$-meson anomalies. We have also approximated $s_{13}^{Ld} \approx - s_{23}^{Ld} s_{12}^{Ld} $, which slightly overestimates the Wilson coefficients, see Eq.~(\ref{eq:approxrotangles}).
These expressions have to be compared to the bounds from $K$-mixing taken from Ref.~\cite{KboundsETM} (Im) and Ref.~\cite{KboundsUTfit} (Re), and  the bounds from $B$-mixing taken from Ref.~\cite{BboundsETM}:

\begin{align}
\label{eq:kaonbounds}
{\rm Im} \, C_1^K & < \frac{3.4 \times 10^{-9}}{\TeV^2} \, ,& {\rm Re} \, C_1^K & < \frac{9.6 \times 10^{-7}}{\TeV^2} \, , \\
|C_1^{B_d}| & < \frac{1.4 \times 10^{-6}}{\TeV^2} \, ,& |C_1^{B_s}| & < \frac{1.8 \times 10^{-5}}{\TeV^2} \, .
\end{align}
This shows that for $s^{Ld}_{23} \sim V_{cb}$ and $k$ in the experimentally preferred range $k \in [1.9,4.9]$ the bounds from $K$, $B$ and $B_s$ mixing are satisfied, even for an $O(1)$ phase in 
${\rm Im} \, C_K^1$.  

Turning to the up sector, 4-fermion operators with both LH and RH fermions are generated, as a result of a non-universal $U(1)_F$ charge assignment. 
In particular, for the $\Delta C = 2$ operators we have (with $m_u/m_c \approx 0.002$)
\begin{align}
C_1^D & = \frac{7.1 \times 10^{-12}}{\TeV^2} 
\left( \frac{s^{Lu}_{23}}{|V_{cb}|} \right)^4
\left ( |V_{cb}| \over  s^{Ld}_{23} \right ) \, k 
  \, , \nonumber\\
\tilde{C}_1^D & =\frac{7.1 \times 10^{-12}}{\TeV^2} 
  \left( \frac{s^{Ru}_{23}}{|V_{cb}|} \right)^4
\left ( |V_{cb}| \over  s^{Ld}_{23} \right ) \, k  
  \, , \nonumber\\
C_5^D & =   \frac{2.8 \times 10^{-11}}{\TeV^2} \left( \frac{s^{Lu}_{23}}{|V_{cb}|} \right)^2 \left( \frac{s^{Ru}_{23}}{|V_{cb}|} \right)^2   
\left ( |V_{cb}| \over  s^{Ld}_{23} \right ) \, k  
 \, . 
\end{align}
Hence, D-meson mixing  is further suppressed  compared to K-meson mixing by the small mass ratio 
$m_u/m_c$. 
Given the bounds from Ref.~\cite{DboundsETM} (Im) and Ref.~\cite{KboundsUTfit} (Abs):
\begin{align}
{\rm Im} \, C_1^D & < \frac{0.9 \times 10^{-8}}{\TeV^2} \, ,& |C_1^D| & < \frac{7.2 \times 10^{-7}}{\TeV^2} \, , \nonumber \\
{\rm Im} \,  \tilde{C}_1^D  & < \frac{0.9 \times 10^{-8}}{\TeV^2} \, ,& |\tilde{C}_1^D| & < \frac{7.2 \times 10^{-7}}{\TeV^2} \, , \nonumber \\
{\rm Im} \, C_5^D & < \frac{0.4 \times 10^{-8}}{\TeV^2} \, ,&|C_5^D| & < \frac{4.8 \times 10^{-7}}{\TeV^2} \, ,  
\end{align}
there are no further bounds on our model  from $D-$mixing.

\subsection{Semileptonic decays in $b \to d$ and $s \to d$ transitions}
We now turn to semileptonic decays involving $b \to d$ and $s \to d$ transitions with electrons or muons in the final state.
The relevant Wilson coefficients for the associated 4-fermion operators for $b \to d$ transitions are given by
\begin{align}
C^{(bd)(ee)} & = \frac{ g'{}^2 \Delta_{L}^{bd}  \Delta_{L}^{ee} }{2 \, M_{Z^\prime}^2} = \frac{3.4 \times 10^{-5}}{\textrm{TeV}^2} \, k \, ,  \nonumber \\
 C^{(bd)(\mu \mu)} & = \frac{ g'{}^2 \Delta_{L}^{bd}  \Delta_{L}^{\mu \mu} }{2 \, M_{Z^\prime}^2} = -\frac{2.3 \times 10^{-5}}{\textrm{TeV}^2} \, k \, , 
\end{align}
and for the $s \to d$ transitions by
\begin{align}
C^{(sd)(ee)} & = \frac{ g'{}^2 \Delta_{L}^{sd}  \Delta_{L}^{ee} }{2 \, M_{Z^\prime}^2} = - \frac{1.4 \times 10^{-6}}{\textrm{TeV}^2}  \left( \frac{s_{23}^{Ld}}{|V_{cb}|} \right) k \, , \nonumber \\ 
C^{(sd)(\mu \mu)} & = \frac{ g'{}^2 \Delta_{L}^{sd}  \Delta_{L}^{\mu \mu} }{2 \, M_{Z^\prime}^2} = \frac{9.4 \times 10^{-7}}{\textrm{TeV}^2}  \left( \frac{s_{23}^{Ld}}{|V_{cb}|} \right) k \, , 
\end{align}
where we approximated $\Delta_L^{bd} \approx - s_{23}^{Ld} \sqrt{m_d/m_s}$ and $\Delta_L^{sd} \approx  \left( s_{23}^{Ld}\right)^2 \sqrt{m_d/m_s}$.
The upper bounds are summarized in the Table~\ref{Tbsbd}, adapted from the case of composite leptoquarks \cite{GripaiosNardecchiaRenner}. 
\begin{table}
\begin{center}
\begingroup
\begin{center}
\begin{tabular}{c c c}
\toprule
Decay & $(ij)(kl)$ & $| \Delta^{d_i d_j}_{L} \Delta^{e_k e_l}_{L} | / (\sqrt{2} \, \frac{M_{Z'}}{1 \textrm{TeV}})^2 $  \\
\midrule
$K_S \to e^+ e^-$ & $ (21)(11)$ &  $< 1.0$ \\
$K_L \to e^+ e^- $ & $(21)(11)$ &  $< 2.7 \times 10^{-3}$  \\
$K_S \to \mu^+ \mu^-$ & $(21)(22)$ &  $< 5.1 \times 10^{-3}$\\
$K_L \to \mu^+ \mu^- $ & $(21)(22)$ &  $< 3.6 \times 10^{-5}$ \\
$K^+ \to \pi^+ e^+ e^-$ & $(21)(11)$ &  $< 6.7 \times 10^{-4}$  \\
$K_L \to \pi^0 e^+ e^- $ & $(21)(11)$ &  $< 1.6 \times 10^{-4}$  \\
$K^+ \to \pi^+ \mu^+ \mu^-$ & $(21)(22)$ &  $< 5.3 \times 10^{-3}$ \\
$B_d \to \mu^+ \mu^-$ & $(31)(22)$ &  $< 3.9 \times 10^{-3}$ \\
$B^+ \to \pi^+ e^+ e^-$ & $(31)(11)$ &  $< 2.8 \times 10^{-4}$ \\
$B^+ \to \pi^+ \mu^+ \mu^-$ & $(31)(22)$ &  $< 2.3 \times 10^{-4}$ \\
\bottomrule
\end{tabular}
\label{BKconstraints}
\end{center}
\endgroup
\end{center}
\caption{\label{Tbsbd} Upper bounds on Wilson coefficients from leptonic and semi-leptonic $K$  and $B$ decays with $s \to d$  and $b \to d$ transitions.}
\end{table}
From this table, it is easy to verify that for the experimentally preferred range $k \in [1.9,4.9]$ all the bounds are satisfied.

\subsection{Lepton flavor violation}

From \eref{zee}, the largest LFV  $Z^\prime$ couplings are the ones to LH muons and electrons. 
These are constrained by several precise measurements of LFV $\mu \to e$ transitions. 
First, we have the $\mu \to 3 e$ decay with the branching fraction 
\begin{align}
{\rm BR} (\mu \to 3 e) & = \frac{g'{}^4 v^4}{4 M_{Z^\prime}^4} 
|\Delta_L^{e \mu}|^2 \left( 2 |\Delta_L^{e e}|^2   +   |\Delta_R^{ee}|^2 \right)  
\nnl 
& \approx 
 3.2 \times 10^{-14}  \, k^2 
\left( \frac{s_{13}^{Le}}{1.1 \times 10^{-4}} \right)^2  
  \left( \frac{|V_{cb}|}{s^{Ld}_{23}} \right)^2 \, , 
\end{align}
where $v = 246$~GeV. 
This should be  compared with the experimental limit from Ref.~\cite{mu3ebound}:
\begin{align}
{\rm BR}  (\mu \to 3 e) & < 1.0 \times 10^{-12} \, .
\end{align}
This limit can be violated for larger values of the parameter  $k$. 
In particular, for $X_1^E  =0 $ and $k$ in the experimentally favored range $k \in [1.9,4.9]$ we get the constraint on the mixing angles: 
\begin{align}
 \left( \frac{|V_{cb}|}{s^{Ld}_{23}} \right)   \left(  \frac{s_{13}^{Le}}{1.1 \times 10^{-4}} \right) < [3.0, 1.2] \, . 
 \end{align}
This can be satisfied for $s_{23}^{Ld} \sim |V_{cb}|$ if the ${\cal O}(1)$ Yukawa coupling controlling  $s_{13}^{Le}$ is $\lesssim 1$.  

A stronger constraint on LFV  comes from $\mu$-$e$ conversion in nuclei. 
Borrowing the formulas  e.g from  Refs.~\cite{Kosmas:2001mv,mueconvHerrero,Porod:2014xia},  
for $X^E_1=0$ the conversion rate in gold, titanium and aluminium nuclei is given by:
\begin{align}
{\rm CR} (\mu \to e, {\rm Au}) & \approx    4.0 \times 10^{-12}   \left( \frac{|V_{cb}|}{s^{Ld}_{23}} \right)^2  \left(  \frac{s_{13}^{Le}}{1.1 \times 10^{-4}} \right)^2    k^2   \, , 
\nnl 
{\rm CR} (\mu \to e, {\rm Ti}) & =   
3.8 \times 10^{-12}   \left( \frac{|V_{cb}|}{s^{Ld}_{23}} \right)^2  \left(  \frac{s_{13}^{Le}}{1.1 \times 10^{-4}} \right)^2    k^2  \, , 
\nnl 
{\rm CR} (\mu \to e, {\rm Al}) & =   
1.9 \times 10^{-12}   \left( \frac{|V_{cb}|}{s^{Ld}_{23}} \right)^2  \left(  \frac{s_{13}^{Le}}{1.1 \times 10^{-4}} \right)^2    k^2  \, ,  
\end{align}
This should be compared with the bounds from Ref.~\cite{mueconvbound,Dohmen:1993mp}: 
\begin{align}
{\rm CR}  (\mu \to e, {\rm Au}) & <  7.0 \times 10^{-13}, 
\quad 
{\rm CR}  (\mu \to e, {\rm Ti})  <  4.3 \times 10^{-12} \, .
\end{align}
For the parameter $k$ in the range favored by the $B$-meson anomalies, $k \in [1.9.4.9]$,  
this leads to the constraint on a combination of mixing angles  in our model 
\begin{align}
\label{eq:conversionconstraint}
 \left( \frac{|V_{cb}|}{s^{Ld}_{23}} \right)   \left(  \frac{s_{13}^{Le}}{1.1 \times 10^{-4}} \right) < [0.22, 0.08] \, . 
 \end{align}
Formally, $s_{13}^{Le}$ is a free parameter, therefore \eref{conversionconstraint} can always be satisfied with an appropriate choice of the lepton Yukawa couplings. 
However, our philosophy is to explain the flavor hierarchies with all Yukawa couplings in \eref{yf} being 
$\cO(1)$, in which case  the natural value is $s_{13}^{Le} \sim 10^{-4}$. In this respect, \eref{conversionconstraint} forces us into a less natural corner of the parameter space and suggest a value of $k$ close to the lower $1\sigma$ boundary.  
We note that the experimental sensitivity to the $\mu$-$e$ conversion rate is expected to improve by many orders of magnitude in the near future \cite{PRIME, COMET, Mu2e}. 
In case of a null result, our model  will no longer be an attractive solution to the $B$-meson anomalies. 

\subsection{Electroweak precision tests}

Integrating out $Z'$ induces lepton-number {\em conserving} 4-fermion operators which can be constrained by electroweak precision tests. 
Here, we focus on the 4-lepton operators which give the strongest bounds due to large $U(1)_F$ charges  of leptons. 
At leading order, these do not affect Z-pole observables measured in LEP-1 and SLC, but they can be constrained by off-Z-pole fermion scattering in LEP-2. 
We parametrize these operators as 
\begin{align}
{\cal L}_{\rm eff} & \supset  
\sum_{\ell \in e,\mu, \tau} \left [   {[c_{LL}]_{e \ell}  \over v^2}  (\bar e \bar \sigma^\mu e)  ( \bar \ell \bar \sigma_\mu \ell ) 
+   {[c_{LR}]_{e\ell}  \over v^2}  (\bar e \bar \sigma^\mu e)  ( \ell^c  \sigma_\mu \bar \ell^c) 
+   {[c_{RR}]_{ef}  \over v^2}  (e^c  \sigma^\mu \bar e^c)  ( \ell^c  \sigma_\mu \bar \ell^c) 
\right ]. 
 \end{align}
 For $X_1^E = 0$ the non-zero Wilson coefficients are 
 \begin{align}
 [c_{LL}]_{ee}  
   &   = -  \frac{g'{}^2 v^2}{2 M_{Z^\prime}^2}  \left ( \Delta_{L}^{e e} \right )^2 
   = - 6.8 \times 10^{-4}  k    \left( \frac{|V_{cb}|}{s^{Ld}_{23}} \right),  
 \end{align}
 \beq
  [c_{LL}]_{e\mu} = - {4 \over 3}   [c_{LL}]_{ee}, \quad 
   [c_{LL}]_{e \tau} = 2  [c_{LL}]_{ee}, \quad 
   [c_{LR}]_{e\tau} = -{8 \over 3} [c_{LL}]_{ee}. 
 \eeq 
 Note that the sign of each contribution is fixed, in particular the contribution to  $[c_{LL}]_{ee}$  is always negative in our model.  
 We calculated the impact of these operators on the LEP-2 observables quoted in Ref.~\cite{LEP2}. 
We used the total cross section and asymmetries of $e^+  e^- \to \mu^+ \mu^-,\tau^+ \tau^-$ 
measured  at the  center-of-mass energies $\sqrt{s} \in [130,207]$~GeV, 
as well as the differential cross-sections of $e^+  e^- \to e^+ e^-$ at  $\sqrt{s} \in [189,207]$~GeV.
This way we obtain the 95\% CL constraint:  
\beq
\label{eq:lep2constraint}
k    \left( \frac{|V_{cb}|}{s^{Ld}_{23}} \right)  \leq   1.1.  
\eeq  
For $k$ in the experimentally favored range $k \in [1.9,4.9]$, 
\eref{lep2constraint} requires somewhat larger values of $s^{Ld}_{23}$, of order $2$-$4$ $|V_{cb}|$ . 
This leads to some tension with the bound from CP violation in kaon mixing in \eref{kaonbounds}, assuming $\cO(1)$ phases entering $C_1^K$. Much as the LFV bound, these constraints point to rather low $k \approx 2$.  
 
We also comment on the corrections to lepton flavor conserving muon decays.  
Loops with a  $Z'$ boson  result in  the following 1-loop correction to the $\mu \to e \nu_\mu \nu_e$ decay width~\cite{CrivellinZ'}: 
\begin{align}
\label{eq:gf1loop}
{\Gamma(\mu \to e \nu \nu) \over \Gamma(\mu \to e \nu \nu)_{\rm SM}}
= 1 -  \epsilon, 
\end{align}
where
\begin{align}
\epsilon & \approx  
- {3 g^{\prime 2} \over 4 \pi^2} \Delta^{ee}_L  \Delta^{\mu \mu}_L {m_W^2 \log(M_{Z'}^2/m_W^2)  \over M_{Z'}^2 } = 1.5 \times 10^{-5} k  \left( \frac{|V_{cb}|}{s^{Ld}_{23}} \right)    \log\left( M_{Z'} \over m_W \right ).  
\end{align}
The muon lifetime measurement does not constrain new physics by itself, because it is used to extract the SM input parameter $G_F$ (equivalently, the Higgs VEV $v$). 
However, indirectly, new physics contributions to $G_F$ shift other observables  (for example $m_W$, Z-pole asymmetries, etc.) away from the SM predictions.
To estimate the resulting constraints, we note that  the effect in \eref{gf1loop} is equivalent to introducing the 4-lepton operator
${[c_{LL}]_{1221} \over v^2 }\bar \ell_1 \bar \sigma_\mu \ell_2 \bar \ell_2 \bar \sigma_\mu \ell_1$, 
with the Wilson coefficient $[c_{LL}]_{1221} = \epsilon$. 
The constraint on this Wilson coefficient from the Z-pole observables can be read off using  the global likelihood function quoted in Ref.~\cite{EfratiFalkowskiSoreq}.   
If only this one operator affects the Z-pole observables, the constraint reads $-0.8 \times 10^{-3} < [c_{LL}]_{1221} < 2 \times 10^{-3}$ at 95\% CL. 
The resulting constraints on the parameter of our model are weaker than the ones from off-Z-pole measurements in LEP-2.


\subsection{$Z^\prime$ searches in colliders}

\begin{figure}[t]
\begin{center}
\includegraphics[scale=0.7]{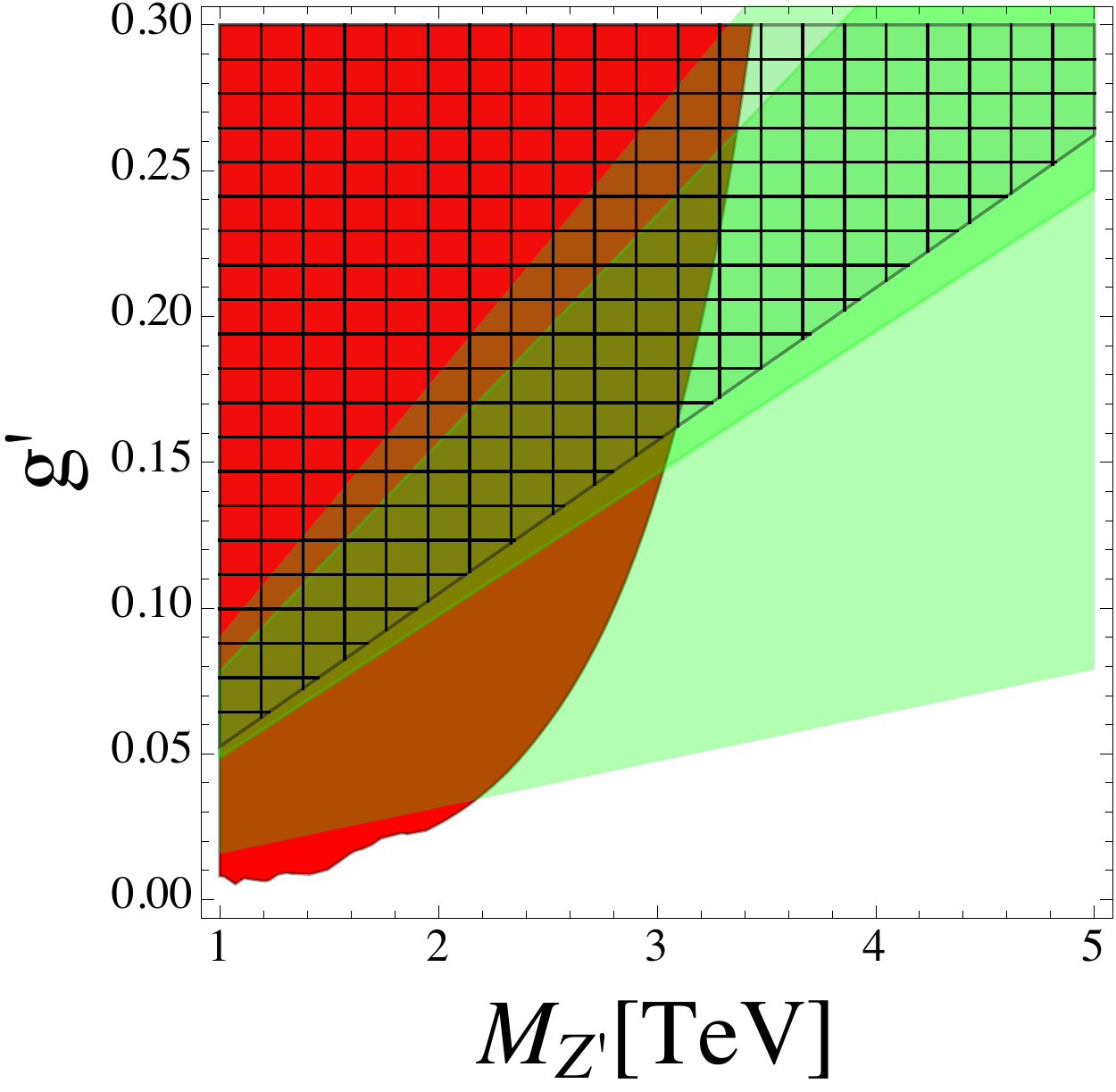}
\caption{
For  $X^E_1=0$, the region of the $M_{Z'}$-$g'$ plane of our model excluded by resonance searches at the LHC (red).  We also show the indirect constraints from the 2-fermion production in LEP-2 (black mesh). 
The green regions correspond to $s_{23}^{Ld} = 2 |V_{cb}|$ and the parameter $k$ in \eref{kdef} 
 in the range favored by the $B$-meson anomalies at  1~$\sigma$  $k \in (1.9,4.9)$ (darker) and at 2~$\sigma$ $k \in (0,2,6.5)$ (lighter).}
\label{fig:direct}
\end{center}
\end{figure}

Finally, the parameter space of our model is constrained by direct searches for resonances  in colliders. 
Since addressing the $B$-meson anomalies requires  $M_{Z'}/g' \sim 20$~TeV, 
the $Z'$ boson predicted by our model is within the kinematic reach of LHC for $g'$ of electroweak strength or smaller. 
Note that the direct searches probe separately the $Z'$ mass and coupling constant, 
unlike all previously discussed observables that depended on these parameters only via  the combination $M_{Z'}/g'$.
Given  the charge assignments in \eref{charges} and \eref{u1f_lepton}, the branching fraction of $Z'$ into dilepton final states is significant.  
In particular, for $X_1^E = 0$,  we have 
\beq
{\rm Br} (Z' \to ee) \approx 14 \%, \quad 
{\rm Br} (Z' \to \mu \mu) \approx 6 \%,
\eeq 
and the strongest constraints are expected from the di-electron channel.
In \fref{direct} we plot the constraints in the $M_{Z'}$-$g'$ plane based on the CMS search for di-electron resonances in the LHC at $\sqrt{s}=8$~TeV   \cite{Khachatryan:2014fba}. 
These constraints imply $M_{Z'} \gtrsim 3$~TeV and $g' \gtrsim 0.1$ in the region of the parameter space  favored by  the  $B$-meson anomalies. 
Note that the direct limits are complementary to the indirect ones from LEP-2. 
The latter would allow us to address the $B$-meson anomalies with a light  ($m_{Z'} \lesssim 2$~TeV) and very weakly coupled $Z'$; such possibility is however excluded by the resonance searches.

\section{Summary and Conclusions}

In this work we have addressed recent anomalies in $b \to s \ell \ell$ transitions. This was achieved by a tree-level exchange of a light $Z^\prime$ vector boson, whose couplings to the SM fermions are governed by an underlying $U(2)_F$ flavor symmetry that explains fermion masses and mixings. While the $U(2)_F$ quantum numbers of quarks are determined by quark masses and CKM angles, there is more freedom in the charged lepton sector. 
However,  requiring that the $b \to s \ell \ell$ anomalies are fit within $1\sigma$ essentially selects unique $U(2)_F$ quantum numbers for charged leptons. The only free parameters are then the $Z^\prime$ mass, the associated gauge coupling and a handful of ${\cal O}(1)$ parameters controlling flavor violation. As a result of the $U(2)_F$ symmetry structure, the magnitude of the latter is set by rotation angles involving the third generation, which implies that flavor violation in the quark sector is strongly suppressed by small CKM angles.


In our model, $Z^\prime$ couples only to LH electrons and muons and has universal couplings to RH down quarks. 
Therefore, the  Wilson coefficients relevant for the $b \to s \ell \ell$ transitions satisfy  $C^{ee,\mu \mu}_{9} =  - C^{ee,\mu \mu}_{10}$, and are therefore approximately aligned with the SM contribution. 
Furthermore, the ratio $C^{ee}_{9}/C^{\mu \mu}_{9}$ is fixed by the $U(2)_F$ quantum numbers. 
As a result,  $b \to s \ell \ell$ transitions are governed by a single parameter $k$ 
\begin{align}
k \equiv 
\left(  \frac{20 \TeV}{M_{Z^\prime}/ g^\prime} \right)^2    \left( \frac{s_{23}^{Ld}}{|V_{cb}|} \right) \, , 
\end{align}
controlling the overall magnitude of the $Z^\prime$ contributions. 
Interference of the new contributions with the SM  is constructive in the electron channel and destructive in the muon channel. 
This leads to the prediction that the semi-leptonic  $B$-meson decay rates  
are rescaled by factors $r_e$ and $r_\mu$  that  are common for all $b\to s ee$ and $b \to s \mu \mu$ processes, respectively. We find
\beq
r_e \approx  (1+0.044 k)^2, \qquad r_\mu \approx  (1-0.029 k)^2  \, , 
\qquad 
R_{X=K,K^*,\phi,X_s} = \left( \frac{1-0.029 k}{1+0.044 k} \right)^2 \, . 
\end{equation}
Using the fits to all available  $B$-meson data from the previous literature, we determined the $1 \sigma$ confidence interval for $k$: 
\begin{align}
k \in [1.9,4.9] \, , 
\end{align}
and thus we predict the  rescaling factors in the electron and muon channels, along with their ratio $R_X$
\begin{align}
r_e  & \in  [1.17,1.48]\, ,  & r_{\mu} & \in  [0.73, 0.89] \, , & R_X & \in [0.50,0.76] \, .
\end{align}
The parameter $k$ also controls flavor-violating processes, but only in conjunction with quark and lepton rotation angles which are fixed up to ${\cal O}(1)$ parameters. The strongest constraints on these parameters arise from $\mu$-$e$ conversion in nuclei. They can be satisfied if the mixing angle $s_{13}^{Le}$ is somewhat suppressed by these $O(1)$ parameters, and/or the mixing angle $s_{23}^{Ld}$ is somewhat enhanced.  These considerations also favor smaller $k$,  close to the lower limit of the  $1\sigma$ confidence interval. Electroweak precision test from LEP-2 lead to similar conclusions. Explicitly, all bounds are satisfied for $k \approx 2$ and the reference values in Eq.~(\ref{eq:benchmark}), for an ${\cal O}(1)$ coefficient $\lesssim  0.5$ in $s_{13}^{Le}$. 

All in all, there exist regions of the parameter space of our model where the $B$-meson anomalies are explained and other experimental  constraints are satisfied. 
The final verdict  will be provided by further of lepton flavor universality in LHCb and B-factories,  as well as by  future experiments looking for $\mu-e$ conversion in nuclei. Apart from indirect searches, the $Z^\prime$ boson is likely within the reach of the LHC run-2, and should first show up in the di-electron channel, as a result of its large coupling to electrons. 

\section*{Acknowledgements}

We thank Gero~v.~Gersdorff, David Straub and Avelino Vicente  for useful discussions.   
AF is supported by the ERC Advanced Grant Higgs@LHC.  This  work was partially made in the ILP LABEX  (under
reference ANR-10-LABX-63) and is supported by
French state funds managed by the ANR within the Investissements d'Avenir programme under reference ANR-
11-IDEX-0004-02. This research was partially supported by the Munich Institute for Astro- and Particle Physics (MIAPP) of the DFG cluster of excellence "Origin and Structure of the Universe". We also thank the Galileo Galilei Institute for Theoretical Physics for hospitality and the INFN for partial support during completion of this work.

\appendix
\numberwithin{equation}{section}
\section{Exact Quark Rotation Angles}   
\setcounter{equation}{0}

In the flavor basis the Yukawa terms read 
\begin{equation}
 {\cal L} \subset +Q^T y_u U H + Q y_d D \tilde{H} + L^T y_e E \tilde{H}\, . 
\end{equation}
where each $y_f$ is a $3 \times 3$ matrix with in general complex elements. 
Here, for simplicity, we assume all entries are real, 
as this will allows us to obtain compact formulas for the eigenvalues and the mixing angles. 
Going to the mass eigenstate basis involves unitary rotations  defined by 
\begin{align}
y_u & = V_L^u \, y^{diag}_u \, (V_R^u)^T  \, , & 
y_d & = V_L^d \, y^{diag}_d \, (V_R^d)^T  \, , &
y_e & = V_L^e \, y^{diag}_e \, (V_R^e)^T  \, , & 
\end{align} 
We parametrize the rotations as 
\be
V_L = V^L_{13} V^L_{12} V^L_{23}, 
\qquad 
V_R = V^R_{13} V^R_{12} V^R_{23}, 
\ee
where ($I=L,R$) and
\begin{align}
V^I_{12} & = 
\begin{pmatrix}
c^I_{12} & s^I_{12} & 0 \\
-s^I_{12} & c^I_{12} & 0 \\
0 &0  & 1
\end{pmatrix} \, , & 
V^I_{13} & = 
\begin{pmatrix}
c^I_{13} & 0 & s^I_{13}  \\
0 & 1  & 0 \\
-s^I_{13} & 0 &  c^I_{13}  
\end{pmatrix}  \, ,  &
V^I_{23} & = 
\begin{pmatrix}
1 & 0  & 0 \\
 0 & c^I_{23}  & s^I_{23}  \\
 0 & -s^I_{23} &  c^I_{23}  
\end{pmatrix}  \, .
\end{align}
We are interested in the Yukawa matrix of the form 
\begin{align}
y & = 
\begin{pmatrix}
0 & y_{12} & 0 \\
- y_{12} & y_{22} & y_{23} \\
0 & y_{32} & y_{33}
\end{pmatrix} \, ,
&
y_{diag} & = 
\begin{pmatrix}
y_1 & 0 & 0 \\
0 & y_2 & 0 \\
0 &0  & 1
\end{pmatrix} \, ,
& y_1 \ll y_2 \ll 1\, .
\end{align}
Plugging this form into the eigenvalue equations, 
from the $1$-$1$, $1$-$3$, $3$-$1$, $1$-$2$, and  $2$-$1$ entries we get 4 independent equations which are
\begin{align}
c_{13}^L c_{13}^R s_{12}^L s_{12}^R \eps_{22} + c_{13}^R  s_{12}^R s_{13}^L \eps_{32} + 
 c_{13}^L s_{12}^L s_{13}^R \eps_{23}  +  s_{13}^L s_{13}^R \eps_{33} + c_{12}^L c_{12}^R c_{13}^L c_{13}^R y_1 & = 0 \, ,\\
c^L_{13} c^R_{13} s_{12}^L \eps_{23}  + c_{13}^R s_{13}^L  \eps_{33} - c_{13}^L s_{12}^L s_{12}^R s_{13}^R \eps_{22}  - 
  s_{12}^R s_{13}^L s_{13}^R \eps_{32} - c_{12}^L c_{12}^R c_{13}^L s_{13}^R y_1 & = 0 \, , \\
c_{13}^L c_{13}^R  s_{12}^R \eps_{32} - c_{13}^R  s_{12
}^L s_{12}^R s_{13}^L \eps_{22} + c_{13}^L  s_{13}^R \eps_{33}- 
  s_{12}^L s_{13}^L s_{13}^R \eps_{23} - c_{12}^L c_{12}^R c_{13}^R s_{13}^L y_1 & = 0 \, , \\
c_{12}^R c_{13}^L s_{12}^L \eps_{22}  + c_{12}^L c_{13}^R s_{12}^R \eps_{22} + c_{12}^R  s_{13}^L \eps_{32} + 
 c_{12}^L s_{13}^R \eps_{23}  - c_{12}^R c_{13}^R s_{12}^L y_1 - c_{12}^L c_{13}^L s_{12}^R y_1 & = 0 \, ,
\end{align}
where we have defined
\begin{align}
\eps_{22} & = s_{23}^L s_{23}^R + c_{23}^L c_{23}^R y_2 \, , 
& 
\eps_{23} & = c_{23}^R s_{23}^L - c_{23}^L s_{23}^R y_2 \, , \\
\eps_{32} & = c_{23}^L s_{23}^R - c_{23}^R s_{23}^L y_2 \, , 
&
\eps_{33} & = c_{23}^L c_{23}^R + s_{23}^L s_{23}^R y_2 \, ,
\end{align}
with
\be
\eps_{22} \eps_{33} - \eps_{23} \eps_{32} = y_2 \, .
\ee
From the first three equations one finds
\begin{align}
s_{12}^R & = - \frac{ s_{13}^R \eps_{33}} {c_{13}^R \eps_{32}} \, , 
&
 s_{12}^L & = - \frac{s_{13}^L \eps_{33} }{c_{13}^L \eps_{23}} \, , 
\label{s12sol}
\end{align}
and 
\be
\label{eqsim}
c_{12}^L c_{12}^R y_1 + \frac{s_{12}^L s_{12}^R y_2}{\eps_{33}} = 0 \, .
\ee
Using this last equation, one gets from the 4th equation the solution
\be
c_{13}^R = \frac{\eps_{33} \left[ (c_{13}^L)^2 \eps_{23}^2 y_1^2 + 
    (s_{13}^L)^2 \left(- \eps_{33}^2 y_1^2 + y_2^2 \right) \right]}{c_{13}^L \eps_{23}^2 y_1 y_2} \,  .
\ee
Using this solution and Eqs.~(\ref{s12sol}) in Eq.~(\ref{eqsim}), one finally can solve for $(c_{13}^L)^2$ in terms of $\eps_{ij}$ and $y_{1,2}$. Although the exact solutions are not very complicated, one can approximate these expressions using $y_1 \ll y_2 \ll 1$ to get the final solution (with a consistent choice of signs):
\begin{align}
s_{13}^L & \approx \sqrt{\frac{\eps_{23}^2 y_1}{\eps_{33} y_2}} &
c_{13}^L & \approx \sqrt{1-\frac{\eps_{23}^2 y_1}{\eps_{33} y_2}}   \\
s_{13}^R & \approx \sqrt{\frac{\eps_{32}^2 y_1}{\eps_{33} y_2}}  &
c_{13}^R & \approx \sqrt{1-\frac{\eps_{32}^2 y_1}{\eps_{33} y_2}} \\
s_{12}^R & = - \frac{ s_{13}^R \eps_{33}} {c_{13}^R \eps_{32}} & 
c_{12}^R & = {\rm sign}(- \eps_{23} \eps_{32}) \sqrt{1- (s_{12}^R)^2} \\
 s_{12}^L & =  - \frac{s_{13}^L \eps_{33} }{c_{13}^L \eps_{23}} & 
 c_{12}^L & = \sqrt{1- (s_{12}^L)^2}. 
\end{align}

\newpage

\bibliographystyle{JHEP} 
 \bibliography{FNZ2}

\end{document}